\documentclass[aps,prl,reprint,groupedaddress,showpacs]{revtex4-1}

\usepackage{graphicx}
\usepackage{url}
\usepackage{color}

\newcommand{\Ree}{R_\mathrm{ee}}

\newcommand{\kB}{k_\mathrm{B}}
\newcommand{\Ns}{N_\mathrm{s}}
\newcommand{\Np}{N_\mathrm{p}}

\begin{document}

\title{Anomalous dynamics of DNA hairpin folding}

\author{R. Frederickx}
\affiliation{KU Leuven, Institute for Theoretical Physics, 
Celestijnenlaan 200D, 3001 Leuven, Belgium}
\author{T. in't Veld}
\affiliation{KU Leuven, Institute for Theoretical Physics, 
Celestijnenlaan 200D, 3001 Leuven, Belgium}
\author{E. Carlon}
\affiliation{KU Leuven, Institute for Theoretical Physics, 
Celestijnenlaan 200D, 3001 Leuven, Belgium}

\date{\today}

\begin{abstract}
By means of computer simulations of a coarse-grained DNA model we
show that the DNA hairpin zippering dynamics is anomalous, i.e.\ the
characteristic time $\tau$ scales non-linearly with $N$, the hairpin
length: $\tau \sim N^\alpha$ with $\alpha >1$. This is in sharp contrast
with the prediction of the zipper model for which $\tau \sim N$. We show
that the anomalous dynamics originates from an increase in the friction
during zippering due to the tension built in the closing strands. From
a simple polymer model we get $\alpha = 1+\nu \approx 1.59$ with $\nu$
the Flory exponent, a result which is in agreement with the simulations.
We discuss transition path times data where such effects
should be detected.
\end{abstract}


\maketitle

The folding dynamics of DNA (or RNA) hairpins, which are single
stranded molecules forming a stem-loop structure, has been a topic
of broad interest within the biophysics community for a long time
\cite{bonn98,shen01,math04a,wood06,chod11,kuzn12,neup12}.  Hairpin folding
is a prototype example of secondary structure formation \cite{math10}
and shares common features with the more complex case of protein folding
\cite{dill12}.  In both cases the folding process is described by a
one-dimensional reaction coordinate performing a diffusive motion across
a free energy potential barrier (see e.g.~\cite{best10}).  Recent advances
in experimental single molecule techniques allow to monitor the folding of
hairpins~\cite{neup12} and of proteins~\cite{chun12} with an unprecedented
time resolution.  These and future experiments are expected to elucidate
many aspects of the folding dynamics~\cite{humm12}, the reason being
that the actual conformational changes occur on timescales which can
be typically a few orders of magnitudes smaller than the total folding
time~\cite{chun12}.  

\begin{figure}[b]
\includegraphics[angle=0,width=0.5\textwidth]{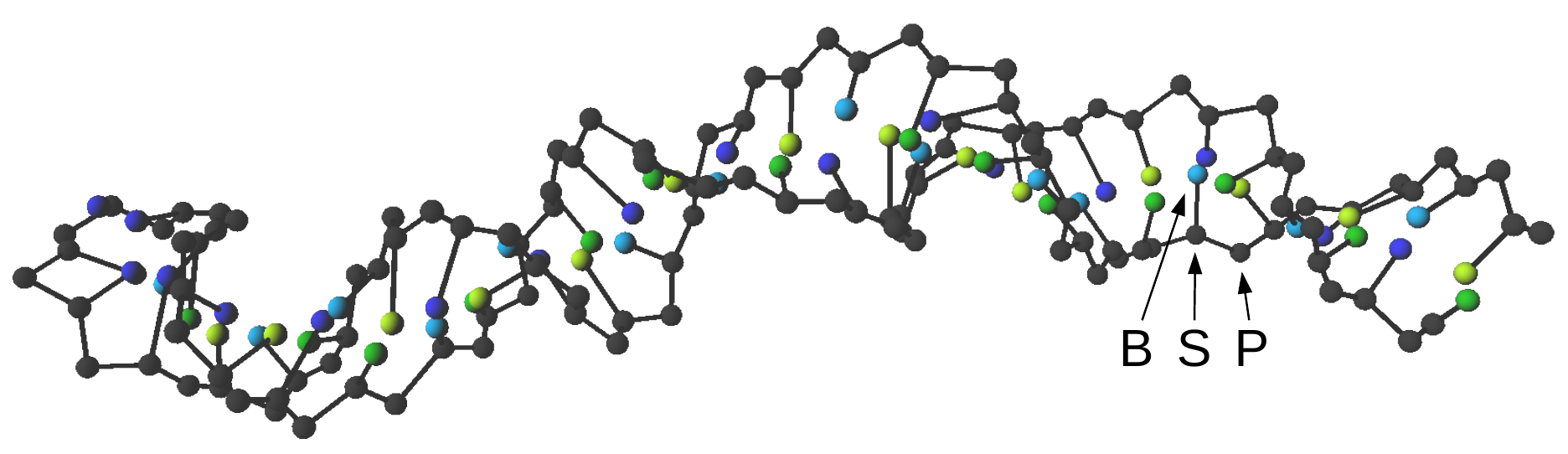}
\caption{Snapshot of a DNA hairpin at the end of the folding process
as simulated by the 3SPN model. The 3 mesoscopic beads are sugar (S),
phosphate (P) and one of the four different bases (A,T,C and G).}
\label{fig01}
\end{figure}

The aim of this letter is to investigate the folding dynamics of DNA
hairpins, focusing in particular on the rapid zippering which follows the
formation of a stable nucleus of a few base pairs. The latter process
is generally much slower as initially the hairpin undergoes a large
number of failed nucleation attempts.  We show here that the zippering
time $\tau$ scales with the hairpin length $N$ as $\tau \sim N^\alpha$
with $\alpha > 1$. This conclusion is based on extensive simulations of
coarse-grained model of DNA and on scaling arguments for polymer dynamics.
Our results are at odds with the zipper model \cite{cocc03} which assumes
that the hairpin closes like a zipper following a biased random walk
dynamics, which implies $\alpha =1$.  The results give insights on the
forces involved in the folding process and in particular in the role
of frictional forces. In addition, as argued at the end of this letter,
recent experiments on transition path times~\cite{neup12} appear to be
better described by a non-linear dependence of zippering time vs. $N$,
supporting the results reported here.

\begin{figure}[t]
\includegraphics[angle=0,width=0.45\textwidth]{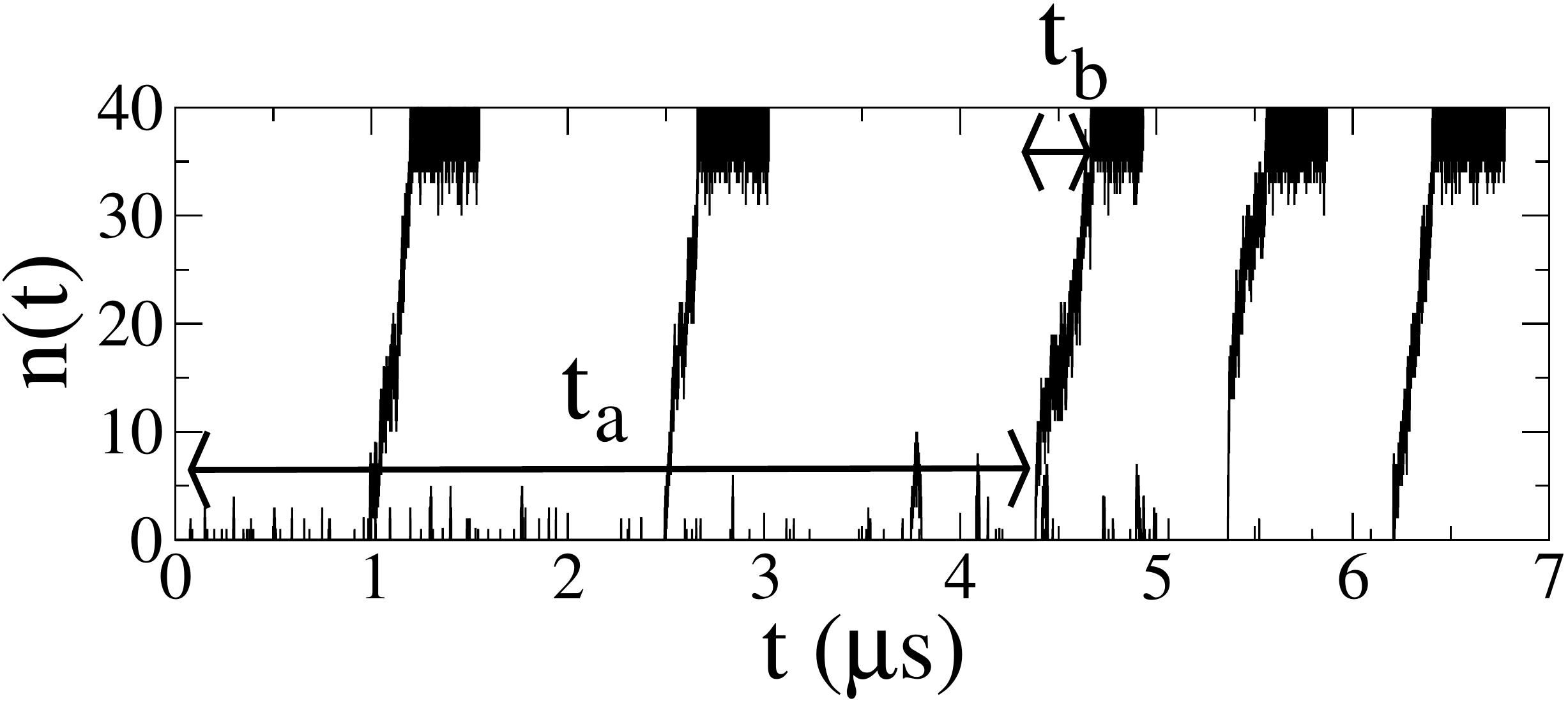}
\caption{Plot of $n(t)$, the number of bound native base pairs,
as a function of time for five different molecular dynamics runs of
folding of an hairpin with stem of length $40$ at $T=10^\circ$C.  In the
simulations only native base pairs interactions are taken into account.
The folding is characterized by a long timescale for the formation of a
nucleus of a few base pairs ($t_{\mathrm a}$), followed by a rapid zippering 
($t_{\mathrm b}$).}
\label{fig02}
\end{figure}

Despite neglecting the fine atomistic details, coarse-grained models
are expected to provide an accurate description of the structure and
dynamics of DNA \cite{knot07,dans10,save10,ould11}. Our simulations
were performed using the three sites per nucleotide (3SPN) model
\cite{knot07}.  Here a nucleotide is mapped to three ``mesoscopic"
beads representing sugar, phosphate and base as shown in the snapshot
of Fig.~\ref{fig01}. The force fields contain interaction terms for
bonds, angles and dihedral angles with equilibrium values reproducing
the B-DNA structure. In addition there are base-pairing, stacking and
electrostatic interactions~\cite{knot07,samb09}. We performed Langevin
dynamics simulations using the BBK integrator \cite{brun84} and with
force fields parametrized as in Ref.~\cite{flor11}. Simulations were
performed at different temperatures $T=10^\circ$C and $T=30^\circ$C and
for hairpins of different lengths with sequences selected as follows. A
single master sequence with a random alternation of AT and CG base pairs was
generated. The hairpin sequences were taken from the master sequence
starting from its origin so that two hairpins of different lengths $N_1 >
N_2$ share the same $N_2$ pairs of nucleotides.

\begin{figure}[t]
\includegraphics[angle=0,width=0.30\textwidth]{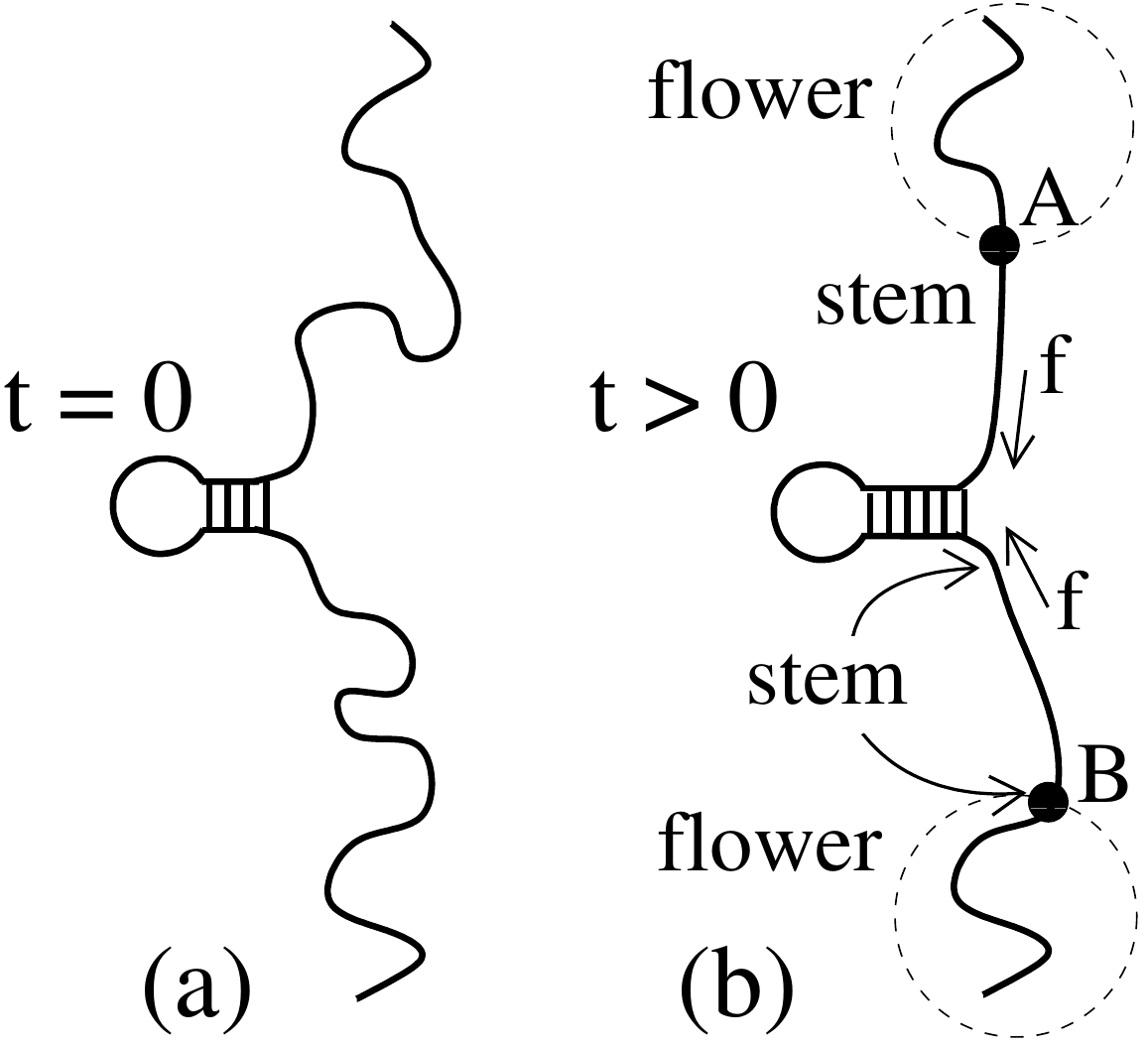}
\caption{(a) Schematic representation of the initial ``clamped"
conformation: a high binding energy is assigned to four base pairs close
to the middle loop so that they remain bound during the simulation
run. (b) During zippering a stem-flower conformation is formed where part
of the single strands are stretched and set into motion (the stems), while
part of the strands are ``unperturbed" as in their original conformation
(the flowers). Here, $f$ denotes the force applied along the backbone
of the strands, close to the fork point.}
\label{fig03}
\end{figure}

As the focus of this paper is the zippering dynamics which follows
the formation of a few native contacts, we consider base-pairing
interactions only between native base pairs, as in the original
3SPN model~\cite{knot07}. Figure~\ref{fig02} shows a plot of $n(t)$
vs.\ $t$, the number of native contacts as a function of time. Two
timescales are visible in the plot: the formation of a stable nucleus
($t_{\mathrm a}$) is followed by a rapid zippering ($t_{\mathrm b}
\ll t_{\mathrm a}$). The analysis of the simulations reveals that the
nucleation predominantly occurs at nucleotides close to the middle of the
strand. Hence, in order to speed up the simulations, we used as initial
state a ``clamped" configuration as that shown in Fig.~\ref{fig03}(a):
a high binding energy was assigned to four pairs of nucleotides close to
the middle of the DNA strand. This energy was chosen sufficiently high so
that the base pairs never unbind during the simulation runs. A single
stranded segment of four adenine nucleotides joins these two clamped
regions together on one side, forming the loop of the hairpin. During
an initial relaxation stage the attractive part of the base pairing
interactions between all the bases in the two strands were turned off,
except for the four clamped base pairs. At a given time ($t=0$) the
attractive energies on the two strands are turned on and the zippering
starts (see Fig.~\ref{fig03}(b)). Note that the repulsive part of the
base pairs interaction is however always on.

\begin{figure}[t]
\includegraphics[angle=0,width=0.45\textwidth]{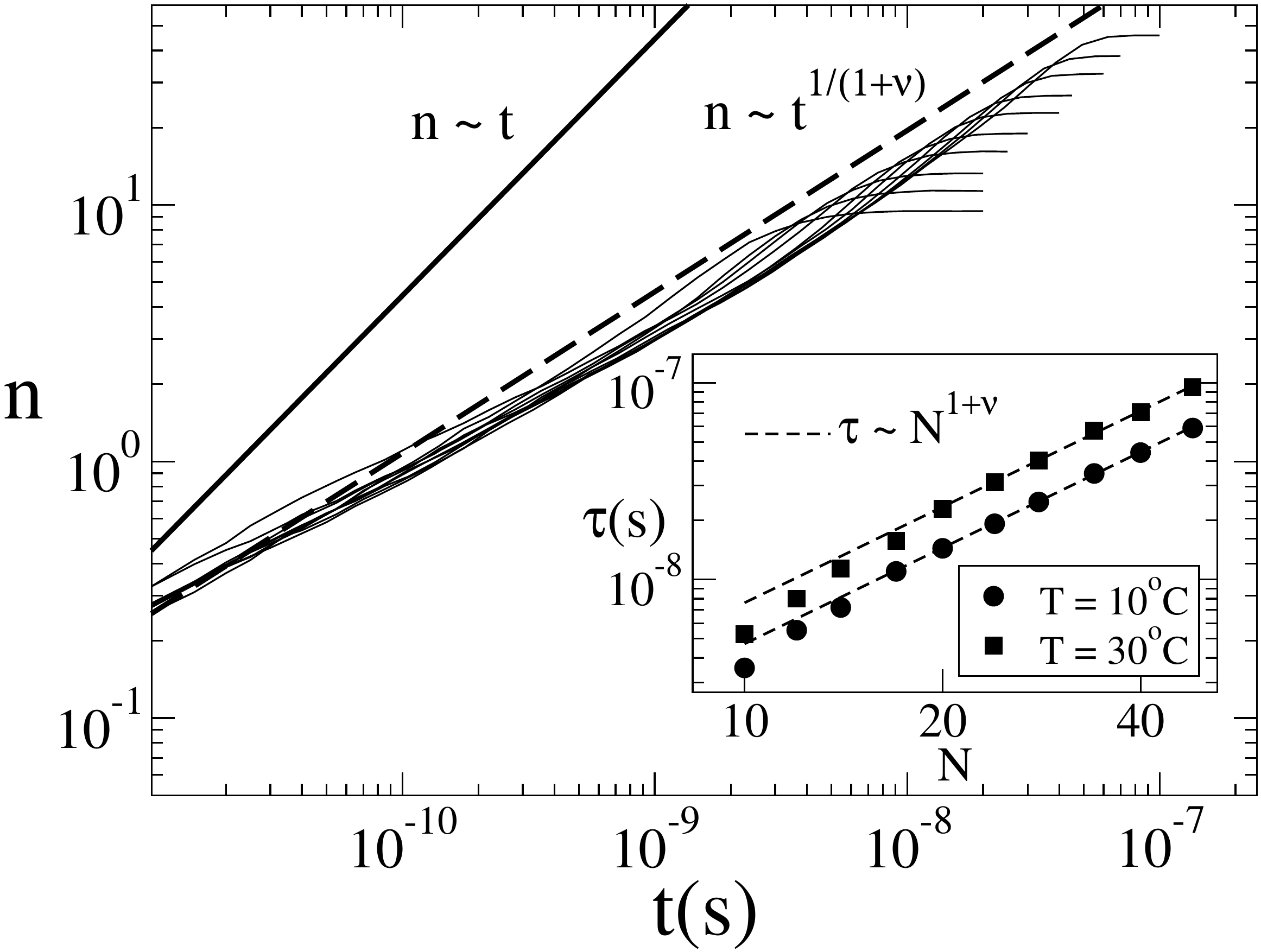}
\caption{Log-log plot of $n(t)$ vs.\ $t$ for various hairpin lengths
$N$ (averaged over typically some 2000 runs) obtained starting from
the clamped conformations of Fig.~\ref{fig03}. The solid line is the
prediction of the zipper model ($n(t) \sim t$), the dashed line is that
from the stem-flower model depicted in Fig.~\ref{fig03} and discussed
in the text. Inset: Log-log plot of zippering times as functions of the
hairpin length.  The two sets of data correspond to hairpin formed at two
different temperatures $T=10^\circ$C and $T=30^\circ$C. The simulation
data are in agreement with the prediction of Eq.~(\ref{langevin}),
shown as dashed lines. Note that there are some finite size effects for
$N < 20$.}
\label{fig:nvst}
\end{figure}

\begin{figure}[t]
\includegraphics[angle=0,width=0.42\textwidth]{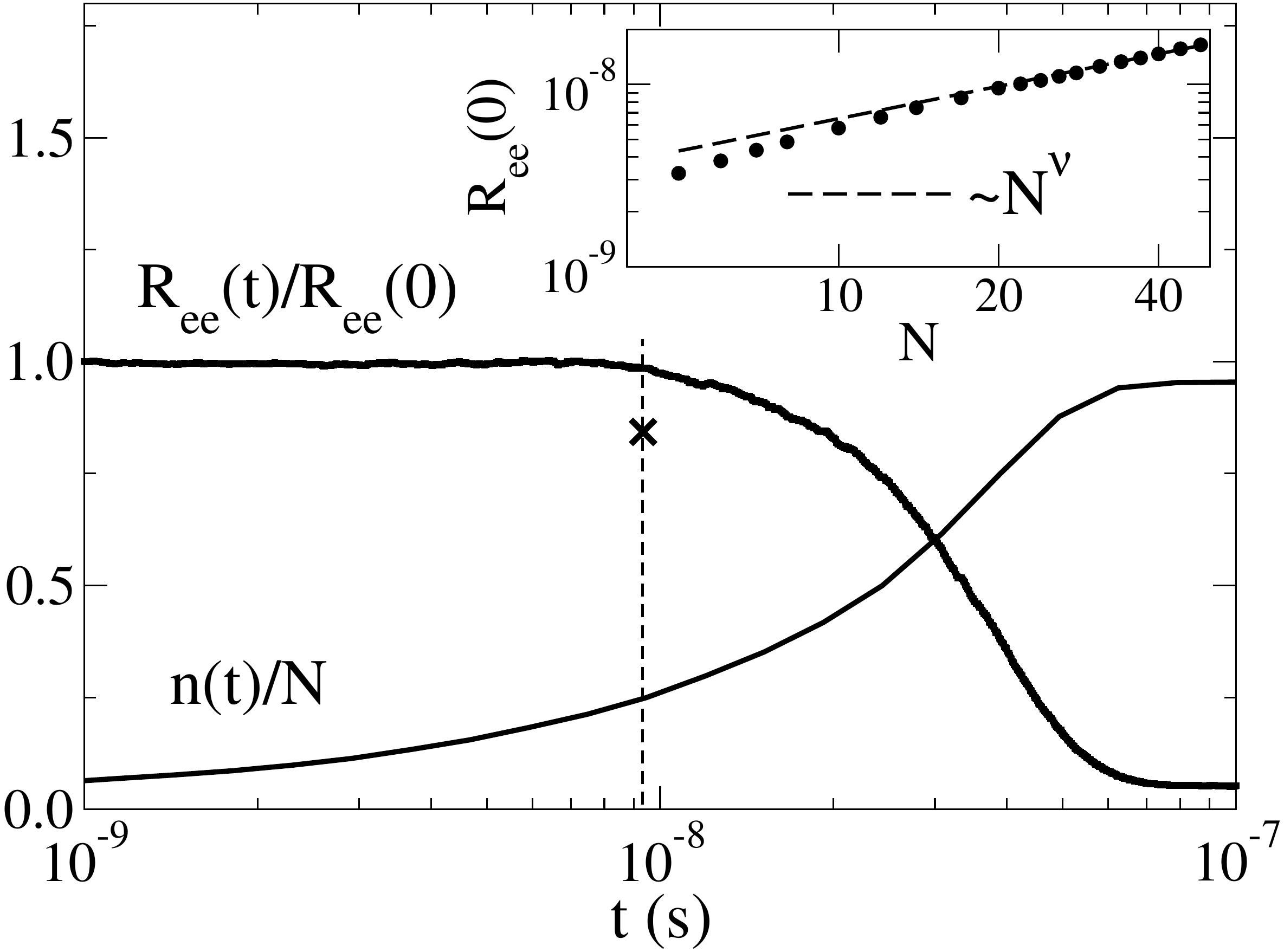}
\caption{Plots of $\Ree(t)/\Ree(0)$ and $n(t)/N$ as functions of time
for the longest hairpin simulated ($N=48$). At the time marked by the
vertical dashed line the end-end distance still retains its initial value
while about $1/4$ of the hairpin bases are formed. The cross indicates
the value of $\Ree(t)/\Ree(0)$ expected if the configuration with
the given $n(t)$ was in equilibrium. Inset: The initial value of $\Ree$
as a function of the hairpin length. The dashed line shows the expected 
asymptotic equilibrium behavior $\Ree(0) \sim N^\nu$.}
\label{fig:Ree_nt}
\end{figure}

Figure~\ref{fig:nvst} shows a plot of the number of formed base
pairs vs.\ time in a log-log scale for a simulation temperature of
$T=30^\circ$C and for hairpins of length $N=10$ to $N=48$. Here $N$
indicates the maximal number of base pairs which can be bound during
the simulation, excluding the initially clamped pairs.  Hence counting
the eight bases which are clamped and four in the loop, a given $N$
corresponds to a sequence of a single strand with $2N+12$ nucleotides.
In Fig.~\ref{fig:nvst} we plot the linear law $n(t)\sim t$ expected in
the zipper model; clearly the dynamics is slower than predicted from
the zipper model. The data instead follow a power-law scaling which
is consistent with $n(t) \sim t^{1/(1+\nu)}$, where $\nu=0.59$ is the
Flory exponent \cite{dege79}. This behavior matches the theory discussed
below.  We estimate the characteristic zippering time by requiring that
the number of formed base pairs is a fraction of the total $N$, i.e.\
$n(\tau) = \lambda N$, where we took different values for $\lambda$ in
the range $0.4 \leq \lambda \leq 0.8$. The inset of Fig.~\ref{fig:nvst}
shows a plot of $\tau$ (circles) obtained by setting $\lambda =0.7$ vs.\
the hairpin length $N$.  The data follow a power-law behavior $\tau \sim
N^\alpha$ with $\alpha = 1.59(2)$ (circles). The simulations were repeated
at $T=30^\circ$C (squares) with a similar result. Taking into account
the results from both temperatures, and the variations arising from the
different possible choices of $\lambda$, we arrive at the aforementioned
final result of $\alpha = 1.60(3)$, which is consistent with $\tau \sim
N^{1+\nu}$.  Figure~\ref{fig:Ree_nt} shows a plot of $\Ree(t)/\Ree(0)$
and $n(t)/N$ vs.\ $t$. The end-end distance $\Ree(t)$ starts from its
maximal value and drops to a small constant value when the hairpin closes,
while $n(t)$ increases as the zippering proceeds. By comparing the two
quantities at equal times (dashed vertical line) one sees that $\Ree$
still largely retains its initial value while roughly a quarter of the
base pairs have already formed. This indicates that only a part of the
single strands are set into motion when the hairpin starts forming,
while the far ends of the two strands are still in their equilibrium
configuration.  Such conformation is known in polymer physics as a
stem-flower shape (see Fig.~\ref{fig03}(b)) and it is the cause of the
anomalous dynamics, as discussed below. The inset of Fig.~\ref{fig:Ree_nt}
shows a plot of the initial value of the $\Ree$ as a function of $N$,
showing that in the 3SPN model the asymptotic regime $\sim N^\nu$ is
reached at around $N=20$ (note that $N$ is the length of a single strand,
hence $\Ree$, the distance between the end points refers to a separation
of $2N$ nucleotides).

The stem-flower dynamics has been discussed in the context of the
absorption of polymers to a flat surface~\cite{desc06}.  The number of
bound base pairs $n(t)$ is expected to follow the equation
\begin{equation} 
\gamma(n) \, \dot{n} = f 
\label{langevin} 
\end{equation} 
where $f$ is the constant force due to base pairing (averaging over
differences between AT and CG base pairs), while the friction $\gamma$
is assumed to be $n$-dependent, since it arises from the stretched stems
whose length varies in time (as the flower remains static, it does not
contribute to the friction). We thus expect $\gamma$ to scale as the
number of bases in the stem: $\gamma \sim \Ns$ \cite{desc06}.  We can
work out the $n$-dependence of the friction coefficient by noticing that
the distance between the static flowers (AB in Fig.~\ref{fig03}) scales
as the end-end separation of a single strand of $2n+2\Ns$ nucleotides
in equilibrium. During zippering, this distance is bridged by the two
stems which are stretched back-to-back yielding a separation $\sim 2\Ns$.
As such~\cite{desc06,bhat08,panj09}:
\begin{equation} 
\gamma (n) \sim N_s \sim \left( n + \Ns \right)^{\nu}
\label{gamma}
\end{equation} 
where we have assumed that the conformation of a strand with $n+\Ns$
nucleotides is described by a self-avoiding walk statistics.  Furthermore
for $n$ sufficiently large we approximate $\gamma (n) \sim n^\nu$
\footnote{We note that $0 \leq n \leq N$, while $0 \leq \Ns \leq N^\nu$,
therefore for large $n$ (and $N$) the approximation is justified. To
estimate the range of $n$ for which this relation is valid we rewrite the
right hand side of Eq.(\ref{gamma}) using the appropriate prefactors as
\begin{equation}
2a \Ns = a \Np \left(\frac{2n+2\Ns}{\Np}\right)^\nu
\label{nsn}
\end{equation}
Here $2a \Ns$ is the distance between the two points AB in
Fig.~\ref{fig03}, $a$ is the distance between two nucleotides and $a \Np$
is the persistence length of single stranded DNA. We solve Eq.~(\ref{nsn})
numerically to get $\Ns$ as a function of $n$, from which we obtain
$\gamma(n) \propto \Ns(n)$.  The single stranded DNA persistence
length is estimated in the range $2 \leq \Np \leq 4$ \cite{knot07}.
Taking $\Np=2$, in the range $20 < n < 50$ the friction is approximated
by a power-law $\gamma(n) \approx n^\delta$ with $\delta \approx 0.54$,
while for $\Np=4$ we get $\delta \approx 0.52$. These values are not
far from the asymptotical value $\nu=0.59$, which suggests that the
range of hairpin simulated the assumption $\gamma (n) \sim n^\nu$ is
a good approximation. We note that for $\Np = 2$ we get for $n=35$ the
value $\Ns\approx 10$, hence the approximation $\gamma (n) \sim n^\nu$
does not require $n \gg \Ns$ to hold strongly.}.
Hence Eq.~(\ref{langevin}) becomes $n^\nu \dot{n} \sim f$
which has solution (with $n(0)=0$):
\begin{equation} 
n(t) \sim t^{1/(1+\nu)} 
\label{nt_theory} 
\end{equation} 
and the total zippering time obtained from $n(\tau) \sim N$ is 
\begin{equation} 
\tau \sim N^{1+\nu}.  
\label{tau_theory} 
\end{equation} 
The theory discussed here is valid in the asymptotic limit of long
DNA strands such that their equilibrium properties are described
by the self-avoiding walks statistics $\Ree \sim N^\nu$. This
behavior is seen in the 3SPN model simulations reported in the inset
Fig.~\ref{fig:Ree_nt}. Although the hairpin simulated are rather short
the data of Fig.~\ref{fig:nvst} show good convergence to the expected
asymptotic behavior.  We note that hydrodynamics interactions do not
modify the predicted exponent in the stem-flower regime, as the friction
originates from the stretched parts of the single strands. Hence the
scaling $\tau \sim N^{1+\nu}$ is expected to be relevant for experiments.

Anomalous dynamics in polymers has been studied a lot in the past decade
(\cite{desc06,bhat08,panj09,saka10,rowg12,ikon13,panj13,ferr11,walt13}). Besides
the already mentioned case of polymer absorption to a planar substrate
\cite{desc06} the exponent $1+\nu$ also governs the dynamics of driven
translocation through a small pore (see e.g.\ \cite{rowg12,ikon13}).
The formation of a stem-flower shape in DNA hairpin dynamics is also
supported by polymer physics arguments. Let us consider a single polymer
pulled by a constant force $f$ applied to one of its end monomers
\cite{broc93}.
A stem-flower conformation arises if the force is large enough
such that~\cite{saka12,rowg12}
\begin{equation}
\Sigma = \frac{f a}{\kB T} \gtrsim 1
\end{equation}
where $a$ is the monomer-monomer distance. In DNA hairpins $f$ is the
force due to base pairing (see Fig.~\ref{fig03}), which can be estimated
from the hybridization free energy per nucleotide: $\Delta G \approx f a$.
For $\Delta G$ we use the experimentally determined values from the
nearest-neighbor model from Ref.~\cite{sant04}. Distinguishing between
weak (AT) and strong (CG) base pairings we obtain estimates $1 \lesssim
\Sigma \lesssim 3.6$ at $T=37^\circ$C and $2 \lesssim \Sigma \lesssim 5$
at $T=10^\circ$C. This suggests that the base-pairing in real DNA hairpin
is sufficiently strong to produce a stem-flower conformation.

\begin{figure}[t]
\includegraphics[angle=0,width=0.48\textwidth]{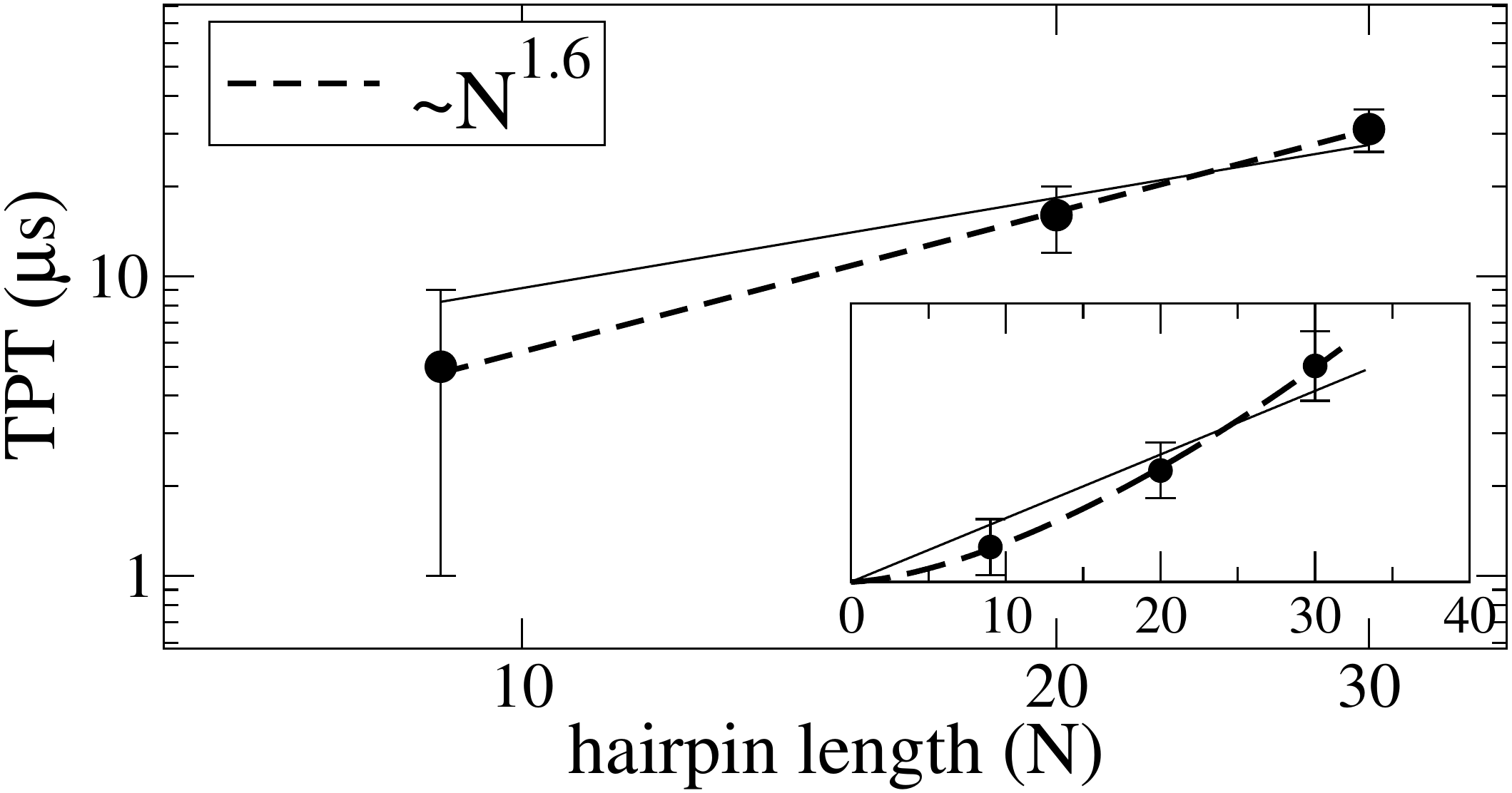}
\caption{Measured transition path times for nucleic acids foldings
as a function of the stem length in a log-log scale (data from
Ref.\cite{neup12}). The results suggest a superlinear scaling, implying
anomalous dynamics. The dashed line is the result of a weighted fit,
giving an exponent $1.6(4)$. The thin solid line is a fitted linear 
law.}
\label{fig:neu_nt}
\end{figure}

There has been quite some recent interest in the experimental
determination of the transition path times (TPT), which are the
short timescales in which the folding process actually takes place
\cite{neup12,chun13}. In analogy to what is shown in Fig.~\ref{fig02}, the
TPT are much shorter than the total folding time and their measurement
is very challenging.  The TPT recently measured in nucleic acids of
different lengths~\cite{neup12} are shown in Fig.~\ref{fig:neu_nt}. We
note that there is a difference in absolute timescales of the simulations
from the 3SPN model of the inset of Fig.~\ref{fig:nvst} and those of
the experiments. This is because the coarse-grained model contains
some simplifications; for instance it does not include explicit solvent
effects~\cite{knot07}, which usually slow down the dynamics. However the
exponent characterizing the dynamical laws is expected to be universal,
despite the difference in absolute times. A weighted fit, which weights
the error bars in each point, of the data of Fig.~\ref{fig:neu_nt} yields
$\tau \sim N^{1.6(4)}$.  The limited data favor a superlinear scaling
compared to a linear scaling as expected from the zippering model.
We note that only very recently TPT have been measured, therefore a
limited amount of data is available. In addition, the TPT are obtained
indirectly as via energy landscape theory~\cite{neup12}, using some
assumptions on the underlying dynamics. It would be interesting to
extend the TPT measurements to test the anomalous dynamics scenario,
which, as shown in this work, is supported by theory and simulations.

Our results show that the simple diffusive motion predicted by the
zipper model cannot explain the simulation data. However, the data are
compatible with a diffusive dynamics with a $n$-dependent diffusion
coefficient, obtained from the fluctuation-dissipation relation $D(n)
= k_BT/\gamma(n)$.  Diffusion coefficients which depend on the reaction
coordinate have been recently discussed in the protein folding literature
\cite{best10}. In the context of the stem-flower folding in DNA hairpin
dynamics the coordinate-dependence arises naturally from the increasing
friction of the closing strands, which should lead to a decrease
in $D(n)$. We expect that this should also happen in the folding of
other biomolecule domains; for instance in the formation of an alpha
helix the two ends of the unfolded polypeptide are pulled towards the
helical domain, producing frictional forces similar to those described
here. Therefore, this discussion could be useful to rationalize the
observed diffusion coefficients in other types of biomolecular folding.


%

\end{document}